\documentclass[aps,prb,twocolumn,nofootinbib, 10pt, superscriptaddress]{revtex4-2}

\usepackage[utf8]{inputenc}
\usepackage{amssymb,amsmath,amsfonts,dsfont,bm}
\usepackage[usenames,dvipsnames]{xcolor}
\usepackage[pdftex]{graphicx}
\usepackage[pdftex,plainpages=false,colorlinks=true,linkcolor=Red, citecolor=blue, urlcolor=blue]{hyperref}

\usepackage{txfonts}

\begin{document}


\title{Fermi arcs \emph{vs} hole pockets: periodization of a cellular two-band model}
\author{S. \surname{Verret}}
\affiliation{D\'epartement de physique \& Institut Quantique, Universit\'e de Sherbrooke, Qu\'ebec, Canada  J1K 2R1}
\author{A. \surname{Foley}}
\affiliation{D\'epartement de physique \& Institut Quantique, Universit\'e de Sherbrooke, Qu\'ebec, Canada  J1K 2R1}
\author{D. \surname{Sénéchal}}
\affiliation{D\'epartement de physique \& Institut Quantique, Universit\'e de Sherbrooke, Qu\'ebec, Canada  J1K 2R1}
\author{A.-M. S. \surname{Tremblay}}
\affiliation{D\'epartement de physique \& Institut Quantique, Universit\'e de Sherbrooke, Qu\'ebec, Canada  J1K 2R1}
\author{M. \surname{Charlebois}}
\email{maxime.charlebois@uqtr.ca}
\affiliation{D\'epartement de physique \& Institut Quantique, Universit\'e de Sherbrooke, Qu\'ebec, Canada  J1K 2R1}
\affiliation{D\'epartement de Chimie, Biochimie et Physique, Institut de Recherche sur l'Hydrog\`ene, Universit\'e du Qu\'ebec a Trois-Rivi\`eres, Trois-Rivi\`eres, Qu\'ebec G9A 5H7, Canada}
\date{\today}
\keywords{}

\begin{abstract}

It is still debated whether the low-doping Fermi surface of cuprates is composed of hole pockets or of disconnected Fermi arcs. Results from cellular dynamical mean field theory (c-DMFT) support the Fermi arcs hypothesis by predicting corresponding Fermi arcs for the Hubbard model. Here, we introduce a simple parametrization of the self-energy, in the spirit of Yang-Rice-Zhang theory, and show that state of the art c-CDMFT calculations cannot give a definitive answer to the question of Fermi arcs vs holes pockets, and this, independently of the periodization (cumulant or Green's function) used to display spectral weights of the infinite lattice. Indeed, when our model is restricted to a cluster and periodized like in c-DMFT, only two adjustable parameters suffice to reproduce the qualitative details of the frequency and momentum dependence of the low energy c-DMFT spectral weight for both periodizations. In other words, even though our starting model has a Fermi surface composed of hole and electron pockets, it leads to Fermi arcs when restricted to a cluster and periodized like in c-DMFT. We provide a new “compact tiling” scheme to recover the hole and electron pockets of our starting non-interacting lattice model, suggesting that better periodization schemes might exist.
\end{abstract}

\maketitle

\begin{figure}
\centering
\includegraphics{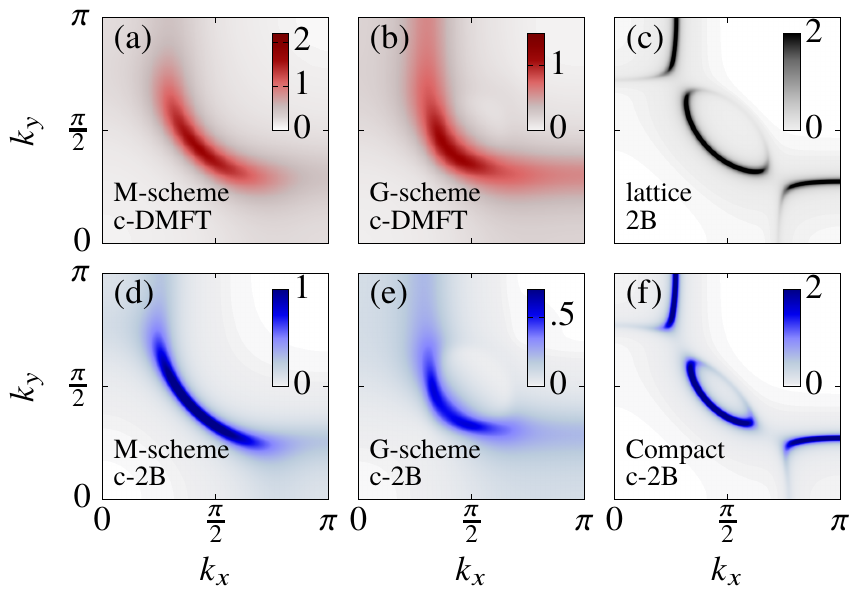}
\caption{
Typical Fermi surface obtained from c-DMFT on hole-doped Hubbard model with (a) the cumulant periodization (M-scheme)~\cite{stanescu_fermi_2006,stanescu_cellular_2006}, and (b) the Green's function periodization (G-scheme)~\cite{kyung_pseudogap_2006,kancharla_anomalous_2008}.
(c) Fermi surface resulting from lattice self-energy~\eqref{lattice_self}, equivalent to an antiferromagnet without broken symmetry.
(d) Fermi arc obtained from the c-2B self-energy~\eqref{eq:sigma1} using the M-scheme~\eqref{cumulant_periodization_main},
(e) the G-scheme~\eqref{green_periodization_main}, and
(f) the compact tiling scheme (proposed at the end of this work).
The band parameters are $t=1$, $t'=-0.3$, $t''=0.2$ and $\eta=0.1$ in all cases. The c-2B parameters $\mu=-0.3$ and $\Delta=0.4$ were chosen to match the c-DMFT Fermi surfaces obtained at interaction $U=8$ and doping $p=0.06$ in the normal state pseudogap (without broken symmetry).
}
\label{figure_fermi_arcs}
\end{figure}

An ongoing debate about the pseudogap state of high temperature superconductors (cuprates) is whether the Fermi surface is composed of small hole-like pockets, as often suggested by transport experiments, or disconnected Fermi arcs, as observed by angle-resolved photoemission spectroscopy (ARPES) experiments~\cite{damascelli_angle-resolved_2003}. Variants of the cluster perturbation theory~\cite{gros_cluster_1993, senechal_spectral_2000}, and more specifically cellular dynamical mean field theory (c-DMFT)~\cite{lichtenstein_antiferromagnetism_2000, kotliar_cellular_2001} provide important theoretical support for the existence of the experimental Fermi arcs: they have found corresponding Fermi arcs in the ground state of the Hubbard model~\cite{senechal_cluster_2002, senechal_hot_2004, kyung_pseudogap_2006, stanescu_fermi_2006,stanescu_cellular_2006, sakai_evolution_2009, sakai_doped_2010}.
However, one important step leading to these theoretical Fermi arcs has long been debated: an operation known as \emph{periodization}~\cite{senechal_cluster_2002, stanescu_fermi_2006, biroli_cluster_2002, biroli_cluster_2004, sakai_cluster-size_2012, verret_intrinsic_2019, klett_real-space_2020}. The main purpose of periodization is to recover a translation-invariant lattice Green's function $G(\mathbf k,\omega)$ from the cluster self-energy $\mathbf \Sigma_c(\omega)$ produced by c-DMFT. This step is necessary only to access the lattice spectral weight and the Fermi surface.
Despite periodization being debated~\cite{sakai_cluster-size_2012}, it is usually accepted that c-DMFT leads to Fermi arcs for the Fermi surface of the pseudogap in the Hubbard model, because the two most widely used periodization schemes: the cumulant periodization (M-scheme)~\cite{stanescu_fermi_2006} and the Green's function periodization (G-scheme)~\cite{senechal_cluster_2002} lead to Fermi arcs, as respectively reproduced in Fig.~\ref{figure_fermi_arcs}(a) and~\ref{figure_fermi_arcs}(b).
However, one question has not been adressed: if the pseudogap of the Hubbard model had a Fermi surface consisting of hole pockets, would c-DMFT correctly capture these hole pockets, given known periodization scheme?

In this letter, we explain how the Fermi arcs obtained with c-DMFT are a consequence of neglecting the self-energy between clusters and using the M- and G-periodization schemes. 
To do so, we start from a non-interacting lattice model leading to a Fermi surface of hole and electron pockets, but here we restrict it to a $2\times2$ cluster. We call this the \emph{cellular two-band} (c-2B) model.
When periodized with the same schemes as in c-DMFT, this minimal model reproduces the spectral weight and associated density of states of the pseudogap state of c-DMFT for the Hubbard model.
This agreement is not limited to the Fermi surface.
Our model reproduces the $\mathbf k$-dependence of the entire low-energy one-particle spectrum obtained in c-DMFT for both the M- and G-schemes. We thus identify multiple $\mathbf k$-dependent structures which are produced by the periodization schemes themselves rather than by many-body correlations. These results bring important clarifications regarding the limitations of current periodization schemes used with cluster methods. We demonstrate that, contrary to previous claims~\cite{stanescu_fermi_2006,stanescu_cellular_2006,sakai_evolution_2009, sakai_doped_2010}, c-DMFT cannot discriminate between Fermi arcs and hole-pockets.

Although our conclusions highlight fundamental limitations of the periodization step within c-DMFT, let us stress that most predictions from c-DMFT regarding the Hubbard model \emph{do not} rely on this step. Notably, c-DMFT allows to study the competition between $d$-wave superconductivity and antiferromagnetism~\cite{capone_competition_2006, kancharla_anomalous_2008, foley_coexistance_2019}, the Mott insulating regime and its interplay with superconductivity~\cite{parcollet_cluster_2004, sakai_evolution_2009, civelli_nodal-antinodal_2008}, the finite temperature transition of the pseudogap~\cite{sordi_finite_2010, fratino_organizing_2016, fratino_pseudogap_2016}, pairing dynamics~\cite{reymbaut_antagonistic_2016}, etc. all of which do not require periodization\footnote{Note also that many of these studies are consistent with the Dynamical Cluster Approximation (DCA), an alternative cluster extension of DMFT. DCA does not require periodization, but does not provide sufficient $\mathbf k$ resolution to obtain the Fermi arcs discussed here.}. These results are out-of-scope for the c-2B model presented here, since the latter cannot self-consistently determine the amplitude of the Mott gap, the chemical potential, the renormalization of enegy scales, the electronic compressibility, the double occupation, etc. The purpose of our c-2B model is to explain the single-particle spectral weight of c-DMFT results, not to capture strongly correlated effects. In that respect,
this work is similar to previous work using density waves models to explain sub-gap structures in the superconducting state of c-DMFT~\cite{verret_intrinsic_2019}, except here we address the normal state pseudogap of c-DMFT.

\paragraph*{Lattice model}
We start from the folowing two-band Hamiltonian on a square lattice (lattice-2B), omitting spin for simplicity,
\begin{align}
H
&=
\sum_{\mathbf k}
\big[ 
\xi^{\phantom\dag}_{\mathbf k} c^{\dagger}_{\mathbf k}c^{\phantom\dag}_{\mathbf k}
+ \xi^{d\phantom\dag}_{\mathbf k} d^{\dagger}_{\mathbf k}d^{\phantom\dag}_{\mathbf k}
+ (\Delta^{\phantom\dag}_{\mathbf k} c^{\dagger}_{\mathbf k}d^{\phantom\dag}_{\mathbf k} + h.c.)
\big]
\label{lattice_Hamiltonian}
\end{align}
Here, $c^{\dagger}_{\mathbf k}$ and $c_{\mathbf k}$ are creation and annihilation operators of electrons with momentum $\mathbf k$ and $d^{\dagger}_{\mathbf k}$ and $d_{\mathbf k}$ are creation and annihilation operators associated to a fermionic auxiliary field. Both $c_{\mathbf k}$ and $d_{\mathbf k}$ are coupled through $\Delta_{\mathbf k}$. 
We do not specify the origin of the auxiliary field in this work.
The electronic dispersion $\xi_{\mathbf k} = \epsilon_{\mathbf k} - \mu$ is given relative to chemical potential~$\mu$ with
\begin{align}
\epsilon_{\mathbf k} &= -2t(\cos k_x + \cos k_y)
- 4t'\cos k_x \cos k_y
\nonumber\\
&\quad
- 2t''(\cos 2k_x + \cos 2k_y),
\label{dispersion_main}
\end{align}
where $t$, $t'$, $t''$ are respectively first- second- and third-neighbor hoppings. The auxiliary field dispersion $\xi^{d}_{\mathbf k} = \epsilon^{d}_{\mathbf k} - \mu^{d}$ is defined equivalently with $t_d$, $t_d'$, $t_d''$.
For the $c^{\dag}_{\mathbf k}$ band, the above lattice model leads to a single-particle Green's function $G(\mathbf k,\omega)=[ \omega +i\eta - \xi_{\mathbf k} - \Sigma(\mathbf k,\omega)]^{-1}$ with real frequency $\omega$, vanishing $\eta\rightarrow0$, and an effective $\mathbf k$-dependent self-energy
\begin{align}
\Sigma(\mathbf k,\omega) =
\dfrac{|\Delta_{\mathbf k}|^{2}}{\omega-\xi^{d}_{\mathbf k}}.
\label{lattice_self}
\end{align}
This self-energy is actually a hybridization function with the $d^\dag_{\mathbf k}$ band.
From the Green's function, the spectral weight is obtained as $A(\mathbf k,\omega)=-2\text{Im}\{G(\mathbf k,\omega)\}$, with Fermi surface $A(\mathbf k,\omega=0)$.

Special cases of the above model include antiferromagnetism (AFM) and the Yang-Rice-Zhang (YRZ) theory, often used to simulate a Fermi surface of hole-pockets and explain transport experiments on cuprates~\cite{yang_phenomenological_2006, leblanc_specific_2009, storey_electron_2013, leblanc_signatures_2014, storey_electron_2013, storey_hall_2016, verret_phenomenological_2017, fang_fermi_2020}.

All results presented in this paper are for a specific case parametrized with $\Delta_{\mathbf k}=\Delta$ and $\xi^{d}_{\mathbf k}=\xi_{\mathbf k + \mathbf Q}$ with $\mathbf Q=(\pi,\pi)$. We choose the values of $\Delta$ and $\mu$ so that this parametrization leads to the same Fermi-surface as an antiferromagnet, with the hole and electron pockets shown in Fig.~\ref{figure_fermi_arcs}(c). 
However, unlike an actual antiferromagnet, Hamiltonian~\eqref{lattice_Hamiltonian} does not break symmetry. This nuance is important since we compare our results to the c-DMFT pseudogap, which does not break symmetry.\footnote{We also verified that YRZ parametrization $\xi^{d}_{\mathbf k}=2t(\cos k_x - \cos k_y)$ with $\Delta_{\mathbf k}=\Delta(\cos k_x + \cos k_y)$ yields equivalent results (not shown).} In the antiferromagnetic phase, 2x2 c-DMFT correctly produces hole pockets.

\paragraph*{Cluster model} We now introduce our cellular two-band (c-2B) model, which is the same model as above, 
but limited to a small cluster with open boundary condition instead of a lattice. 
The effective cluster self-energy is then
\begin{align}
\mathbf \Sigma_c^{\text{c-2B}}(\omega)
&=
\mathbf \Delta^\dag_c \dfrac{1}{\omega-\mathbf t^d_c} \mathbf \Delta_c,
\label{eq:sigma1}
\end{align}
where the coupling $\mathbf\Delta$ and hopping $\mathbf t_c$ are now matrices in the cluster sites positions $\mathbf R$ and $\mathbf  R'$ (see Appendix~\ref{complete}). 

Here we consider the case of a $2\times2$ square cluster in order to compare to $2\times2$ c-DMFT results. With positions enumerated anti-clockwise, we use hopping and coupling matrices
\begin{align}
\mathbf t^d_c
&=
\begin{pmatrix}
 -\mu &  t   &  -t'  &  t  \\
 t   &  -\mu &  t   &  -t' \\
 -t'  &  t   &  -\mu &  t  \\
 t   &  -t'  &  t   &  -\mu
\end{pmatrix}
,\qquad
\mathbf \Delta_c
=
\begin{pmatrix}
\Delta&0&0&0 \\
0&\Delta&0&0 \\
0&0&\Delta&0 \\
0&0&0&\Delta
\end{pmatrix}.
\label{tdc}
\end{align}
This model (c-2B) is equivalent to the two-band lattice model (lattice-2B, Eq.~\eqref{lattice_Hamiltonian}) leading to the hole and electron pockets of Fig.~\ref{figure_fermi_arcs}(c), but restricted to a $2\times2$ cluster.\footnote{The cellular version of YRZ theory (c-YRZ) is obtained with
\begin{align*}
\mathbf t^d_c=&
\begin{pmatrix}
 0 &  t   &  0  &  t  \\
 t   &  0 &  t   &  0 \\
 0  &  t   &  0 &  t  \\
 t   &  0  &  t   &  0
\end{pmatrix}
,\qquad
\mathbf \Delta_c=
\begin{pmatrix}
0&\Delta&0&-\Delta \\
\Delta&0&-\Delta&0 \\
0&-\Delta&0&\Delta \\
-\Delta&0&\Delta&0
\end{pmatrix}.
\end{align*}
Note that YRZ theory usually includes Gutzwiller factors that modify the band structure as a function of doping, but using $(t, t',t'')=(1, -0.35, 0)$ approximate these effects reasonably well for the dopings studied here.
}

\begin{figure*}
\centering
\includegraphics{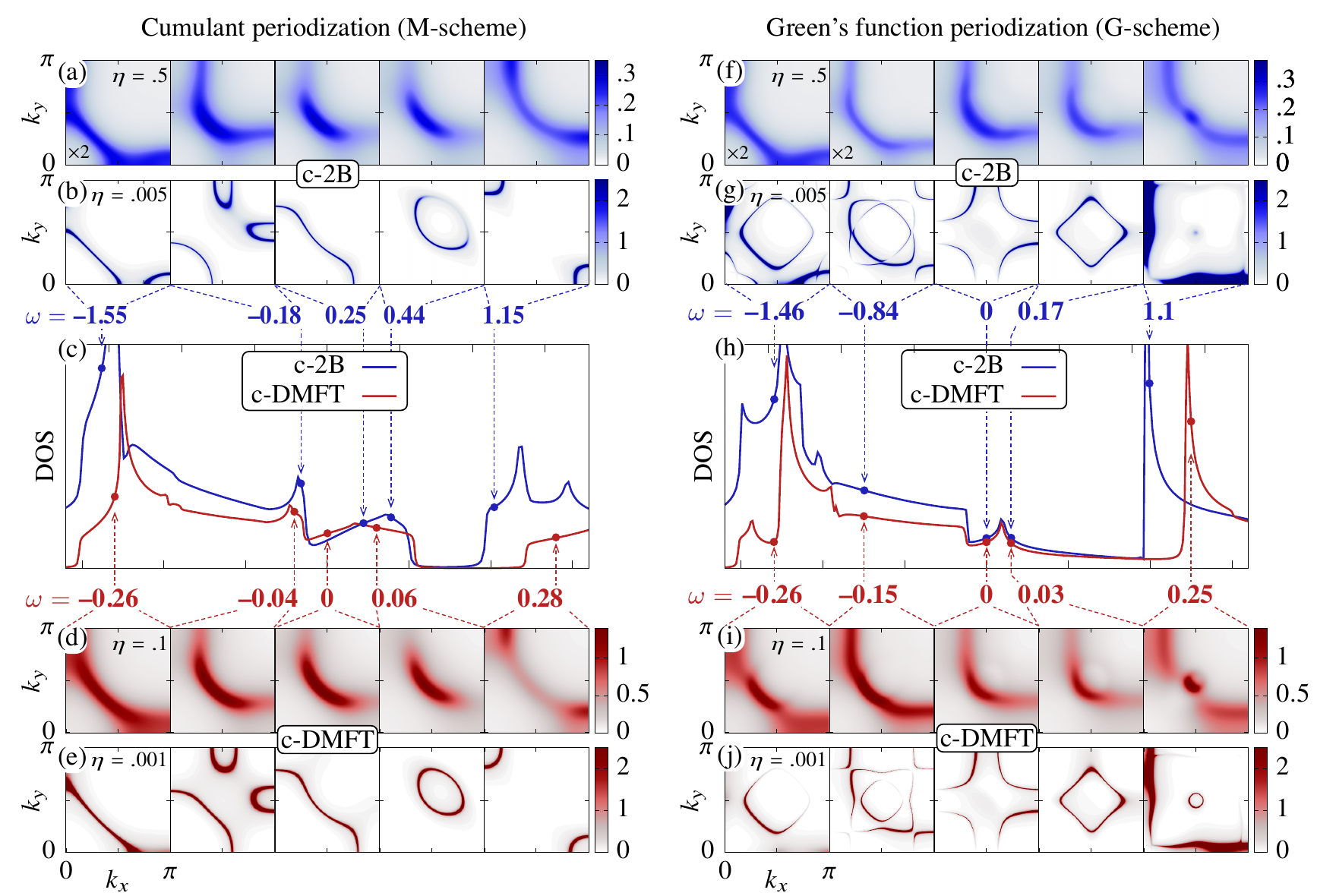}
\caption{
Spectral weight of the c-2B model with $\mu=-0.85$ and $\Delta=1.1$ (blue) obtained with (a) the M-scheme periodization for $\eta = 0.5$ and (b) $\eta = 0.005$. The frequencies shown are indicated along (c) the corresponding density of states (DOS) for $\eta=0.005$. The latter is juxtaposed to the c-DMFT DOS obtained from the M-scheme periodization (red) for $\eta=0.001$. The corresponding c-DMFT spectral weight is shown at other indicated frequencies for (d) $\eta = 0.1$ and (e) $\eta = 0.001$. A factor of 0.18 between two energy axis justifies using different $\eta$. (f-j) shows the same comparison between c-2B results and c-DMFT results, but for the G-scheme periodization.
}
\label{figure_energies_G}
\end{figure*}

\paragraph*{c-DMFT}
We compare the above model with standard results from c-DMFT~\cite{capone_competition_2006,stanescu_cellular_2006,stanescu_fermi_2006, kyung_pseudogap_2006, kancharla_anomalous_2008,civelli_nodal-antinodal_2008,civelli_nodal-antinodal_2008,sakai_evolution_2009,sakai_doped_2010,sakai_cluster-size_2012}. Our implementation uses an exact diagonalization impurity solver of a $2\times2$ cluster with local interactions ($U=8$) and 8 non-interacting baths. Details of our implementation are described in Ref.~\cite{foley_coexistance_2019}, under the name ``simple bath parametrization'', and leads to state-of-the-art results~\cite{senechal_bath_2010, foley_coexistance_2019,kancharla_anomalous_2008,senechal_introduction_2008, kyung_pairing_2009, okamoto_dynamical_2010,senechal_resilience_2013, simard_superfluid_2009,faye_pseudogap--metal_2017, lu_proximity_2020, lu_spin_2021}. The band-structure used for our c-DMFT calculations $(t,t',t'')=(1,-0.3,0.2)$ is the same as in our c-2B model, and hole-doping is chosen at $p=0.06$; in the pseudogap regime. 
We do not allow for broken symmetries such as antiferromagnetism and superconductivity, in order to study the normal state pseudogap. 
In the end, c-DMFT yields a cluster self-energy $\mathbf \Sigma^{\text{c-DMFT}}_c(\omega)$ expressed in the same basis as $\mathbf \Sigma^{\text{c-2B}}_c(\omega)$.

\paragraph*{Periodization} Recovering a translation-invariant lattice Green's function $G(\mathbf k, \omega)$ from a cluster self-energy $\mathbf\Sigma_{c}(\omega)$ is a task known as periodization. The two most widely used periodization schemes for c-DMFT are the M-scheme, or cumulant periodization and the G-scheme, or Green's function periodization.

\paragraph*{M-scheme}
With the cumulant defined as $\mathbf M_c(\omega) = [\omega + i\eta +\mu - \mathbf\Sigma_{c}(\omega)]^{-1}$, the cumulant periodization~\cite{stanescu_strong_2004, stanescu_fermi_2006} is:
\begin{align}
M(\mathbf k, \omega)
=
\frac{1}{N_c}\sum_{\mathbf R,\mathbf R'} e^{-i\mathbf k(\mathbf R - \mathbf R')}
\left[
\frac{1}{\omega+i\eta+\mu - \mathbf\Sigma_{c}(\omega)}
\right]_{\mathbf R,\mathbf R'}
\label{cumulant_periodization_main}
\end{align}
where $N_c=4$ is the size of the cluster. From there, the M-scheme Green's function is obtained as $G^{\text{M}}(\mathbf k,\omega) = [M(\mathbf k, \omega)^{-1} - \epsilon(\mathbf k)]^{-1}$. Fig.~\ref{figure_fermi_arcs}(a) and~\ref{figure_fermi_arcs}(d) show the corresponding Fermi surfaces which consist of Fermi arcs, both for c-DMFT ($\mathbf \Sigma_c \rightarrow \mathbf \Sigma^{\text{c-DMFT}}_c$) and for our c-2B model ($\mathbf \Sigma_c \rightarrow \mathbf \Sigma^{\text{c-2B}}_c$).

\paragraph*{G-scheme}
The Green's function periodization~\cite{senechal_cluster_2002, senechal_quantum_2015} is:
\begin{align}
G^{\text{G}}(\mathbf k,\omega) 
= \frac{1}{N_c}\sum_{\mathbf R,\mathbf R'} e^{-i\mathbf k(\mathbf R - \mathbf R')}
\left[\frac{1}{\omega+i\eta - \mathbf t({\mathbf k}) - \mathbf \Sigma_{c}(\omega)}\right]_{\mathbf R,\mathbf R'}.
\label{green_periodization_main}
\end{align}
This is equivalent to building a super-lattice of clusters connected by the inter-cluster hoppings contained in the lattice dispersion $\xi_{\mathbf k}$~\cite{senechal_spectral_2000, senechal_cluster_2002}. The hopping matrix $\mathbf t(\mathbf k)$ is the cluster representation of $\xi_{\mathbf k}$, with elements given by $t_{\mathbf R, \mathbf R'} (\mathbf k)=\sum_{\mathbf a} e^{-i\mathbf k\cdot\mathbf a} t_{\mathbf a + \mathbf R, \mathbf R'}$ and $\mathbf a$ spanning all super-lattice vectors (see Appendix~\ref{complete}). Fig.~\ref{figure_fermi_arcs}(b) and~\ref{figure_fermi_arcs}(e) show the G-scheme Fermi surfaces, which also consist of Fermi arcs, both for c-DMFT ($\mathbf \Sigma_c \rightarrow \mathbf \Sigma^{\text{c-DMFT}}_c$) and for our c-2B model ($\mathbf \Sigma_c \rightarrow \mathbf \Sigma^{\text{c-2B}}_c$).

\paragraph*{Finite frequency}
The correspondence between the spectral weights of c-2B and c-DMFT for both periodization schemes is striking and not limited to the Fermi level; it extends in the low-energy regime. By fixing $\mu$ and $\Delta$, and with slight adjustments of frequencies, we get a c-2B spectral weight for which the momentum dependence maps almost perfectly to that of c-DMFT. This mapping can be seen in Fig.~\ref{figure_energies_G} for both periodization schemes. For the M-scheme, the c-2B and c-DMFT spectral weights are compared at large $\eta$ in Fig.~\ref{figure_energies_G}(a) and~\ref{figure_energies_G}(d) and at small $\eta$ in Fig.~\ref{figure_energies_G}(b) and~\ref{figure_energies_G}(e).
For the G-scheme, they are compared at large $\eta$ in Fig.~\ref{figure_energies_G}(f) and~\ref{figure_energies_G}(i), and at small $\eta$ in Fig.~\ref{figure_energies_G}(g) and~\ref{figure_energies_G}(j).
Although such small $\eta$ is rarely used in the litterature, 
it unveils the subtle structures contained in the c-DMFT spectral weight.
We observe that these structures are correctly reproduced by our c-2B model.
The frequencies at which matching spectral weights are found are indicated in Fig.~\ref{figure_energies_G}(c) and~\ref{figure_energies_G}(h) along their respective density of states, $N(\omega) = \int d^2k A(\mathbf k, \omega)$. Comparing the density of states further reveals that the sequence of $\mathbf k$-dependent structures observed as a function of $\omega$ is the same in c-DMFT and c-2B.
As expected from a comparison between a non-interacting model (c\nobreakdash-AF) and a strongly-correlated method (c\nobreakdash-DMFT), this qualitative agreement is not perfect; notably, the energy scales are renormalized by a factor of approximately 0.18 between the two models.
Note that the frequencies shown in Fig.~\ref{figure_energies_G} were chosen because of the complexity of the $\mathbf k$-structures they present at very low $\eta$, but equally good fits can be found for all frequencies.
Note also that the parameters $\mu$ and $M$ in Figs.~\ref{figure_fermi_arcs} and~\ref{figure_energies_G} are different. Fig.~\ref{figure_fermi_arcs} highlights that Fermi arcs can be obtained from a system which should have hole and electron pockets, whereas Fig.~\ref{figure_energies_G} focuses on the agreement of the periodized spectral features between c-DMFT and c-2B.

The full mathematical details of the c-2B model are described in Appendix~\ref{complete}
and the source code of the program used to produce the c-2B results of Figs.~\ref{figure_fermi_arcs} and~\ref{figure_fermi_arcs}
can be found in supplemental matrial.

\paragraph*{Discussion}

It is remarkable that identical sequences of structures are obtained with c-DMFT and the c-2B model with both periodization schemes, especially considering that the c-2B model has only two adjustable parameters $\mu$ and $\Delta$. 
This indicates that neglecting the self-energy between clusters and periodizing with the M- and G-schemes is what causes these complex structures, not the strongly correlated physics. 
Moreover, it is these complex structures at small~$\eta$, that blur with those of neigboring frequencies to form the Fermi arcs at large~$\eta$.

Despite these caveats, the Fermi arcs of these methods might have a physical interpretation.
They are obtained at larger values of $\eta$, which can be interpreted as a large electronic scattering rate, and although the G-scheme does not restore the spectral weight of our starting model, it does produce the correct spectral weight for a super-lattice of disconnected cluster self-energies (see section II D of Ref.~\onlinecite{verret_intrinsic_2019}). Therefore, high scattering rate (high $\eta$) and broken translation invariance (disconnected self-energies) seem to be key ingredients leading to Fermi arcs in c-DMFT and our c-2B model. Yet, for these arcs to manifest, a primordial pseudogap mechanism must already be in place, as embodied by $\Delta$ in our c-2B model.


It is important to stress that, although the Fermi-surface of our starting lattice model is the same as that of an antiferromagnet, our results do \emph{not} imply that the c-DMFT pseudogap is a cellularized antiferromagnet. Other parametrization give very similar results. For example, we verified that a YRZ parametrization yields qualitatively equivalent results (not shown) to those presented here.
Moreover, multiple lattice models could match the c-DMFT results, as it seems that only a few poles (the eigenvalues of $\mathbf t_c^d$) in the cluster self-energy are necessary to get this agreement.
This is why our starting lattice model was left general.

The one thing we can conclude from our results is that with the current periodization schemes, $2\times2$ c-DMFT is not equipped to discriminate between a Fermi surface consisting of Fermi arcs and one consisting of hole and electron pockets.
Indeed, the question that guided our work is: ``if the pseudogap of the Hubbard model had a Fermi surface consisting of hole pockets, would c-DMFT correctly capture these hole pockets, given known periodization schemes?'' and the answer is no. As we showed, the M- and G-schemes, arguably the most reliable periodizations~\cite{sakai_cluster-size_2012}, both produce Fermi arcs (Fig.~\ref{figure_fermi_arcs}(d) and~\ref{figure_fermi_arcs}(e)) even when hole-pockets are built in the lattice model from which the cluster model is derived (Fig.~\ref{figure_fermi_arcs}(c)).

\begin{figure}
\centering
\includegraphics[width=0.8\columnwidth]{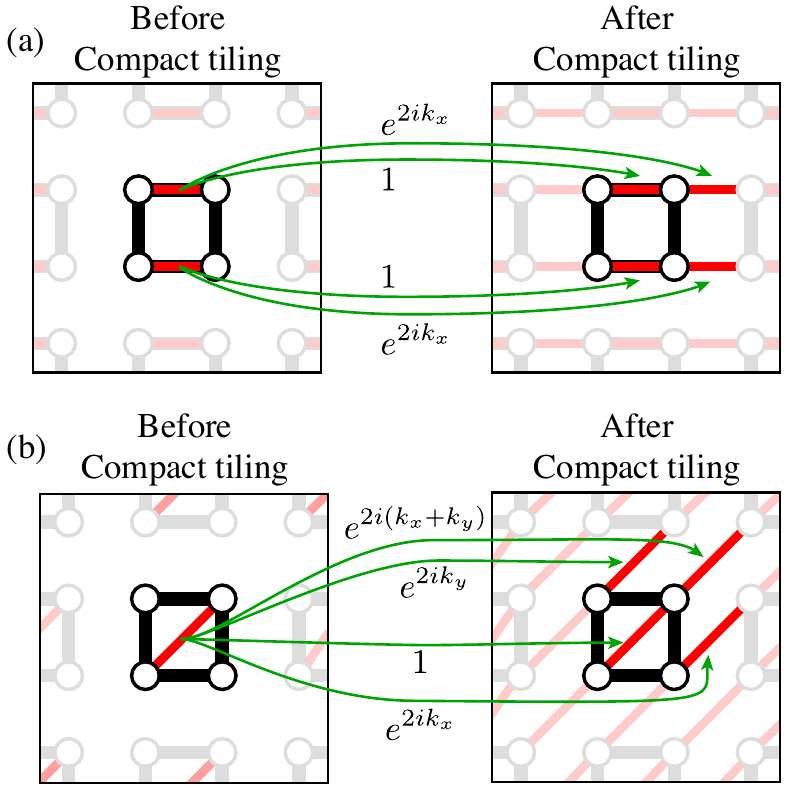}
\caption{Example of compact tiling for the components of $\mathbf t^d_c$ and $\mathbf \Delta_c$. (a) The $\mathbf k$-dependence of matrix components with $\mathbf R-\mathbf R' = (1,0)$ is obtained as $[\mathbf t^d_c(\mathbf k)]_{\mathbf R, \mathbf R'} \approx (1 + e^{2ik_x})[\mathbf t^d_c]_{\mathbf R, \mathbf R'}$, (b) that of components with $\mathbf R-\mathbf R' = (1,1)$ is obtained as $[\mathbf t^d_c(\mathbf k)]_{\mathbf R, \mathbf R'} \approx (1 + e^{2ik_x} + e^{2ik_y} + e^{2i(k_x+k_y)})[\mathbf t^d_c]_{\mathbf R, \mathbf R'}$, and so on for all components of $\mathbf t^d_c$ and with identical expressions for the components of $\mathbf \Delta_c$. This operation results in new \emph{matrices} $\mathbf t^d_c(\mathbf k)$ and $\mathbf \Delta_c(\mathbf k)$ that are invariant under translation.}
\label{fig_compact_tiling}
\end{figure}

Finally, although the correct periodization for c-DMFT is an open question, the right periodization to recover our starting lattice model is quite obvious. What is lacking from cluster self-energy~\eqref{eq:sigma1} to recover the lattice self-energy~\eqref{lattice_self} are the inter-cluster coupling $\delta \mathbf \Delta (\mathbf k) = \mathbf \Delta (\mathbf k) - \mathbf \Delta_c$ and inter-cluster hoppings $\delta\mathbf t^d (\mathbf k) = \mathbf t^d (\mathbf k)- \mathbf t^d_c$ (see Appendix~\ref{complete}). These inter-cluster quantities can be estimated from $\mathbf \Sigma^{\text{c-2B}}_c(\omega)$, given a way to acess $\mathbf t^d_c$ and $\mathbf \Delta_c$ separately.
In such case, we can ``tile'' each element of $\mathbf t^d_c$ and $\mathbf \Delta_c$ to fill the space between clusters as described in Fig.~\ref{fig_compact_tiling}. We call this operation \emph{compact tiling}. Note that we cannot recover $\delta\mathbf t^d (\tilde{\mathbf k})$ perfectly, because $t''$ is absent from $\mathbf t^d_c$.
The results of compact tiling are new $\mathbf k$-dependent matrices $\mathbf t^d_c(\mathbf k)$ and $\mathbf \Delta_c(\mathbf k)$ that allow to define a $\mathbf k$-dependent self-energy matrix $\mathbf \Sigma^{\text{c-2B}}_c(\mathbf k) = \mathbf \Delta_c(\mathbf k)^\dag [\omega - \mathbf t^d_c(\mathbf k)]^{-1}\mathbf\Delta_c(\mathbf k)$. This self-energy can then be used in the M- and G-periodization schemes (setting $\mathbf \Sigma_c \rightarrow \mathbf \Sigma^{\text{c-2B}}_c(\mathbf k)$ in~\eqref{cumulant_periodization_main} and~\eqref{green_periodization_main}). Interestingly, when applied to $\mathbf \Sigma^{\text{c-2B}}_c(\mathbf k)$ (already translation invariant), these periodization formulae amount to simple unitary transformations, and both the M- and G-schemes lead to the same spectral weight.
The corresponding Fermi surface, illustrated in Fig.~\ref{figure_fermi_arcs}(f), presents hole and electron pockets similar to those of the starting lattice model in~Fig.~\ref{figure_fermi_arcs}(c). The only differences are due to $t''$ being absent from the compactly tiled self-energy.
Note, however, that although this approach is simple when applied to our c-2B model, it is not clear yet how it could be generalized to c-DMFT.
Nevertheless, it suggests that better solutions than the M- and G-schemes might exist to perform periodization.

\begin{acknowledgments}

    This research was undertaken thanks in part to funding from the Canada First Research Excellence Fund, the Natural Sciences and Engineering Research Council (Canada) under Grant Nos. RGPIN-2014-04584, RGPIN-2015-05598 and RGPIN-2019-05312, the Research Chair in the Theory of Quantum Materials and postdoctoral research scholarship (B3X) of Fonds de recherche du Québec. Computing resources were provided by Compute Canada and Calcul Québec. 

\end{acknowledgments}

\appendix
\section{Complete derivation}
\label{complete}

\paragraph*{Lattice}

We consider a lattice of $N$ sites with the following definition for the Fourier transforms
\begin{align}
c_{\mathbf r} = \frac{1}{\sqrt{N}}\sum_{\mathbf k}e^{i\mathbf k\cdot\mathbf r}c_{\mathbf k},
\quad
c_{\mathbf k} = \frac{1}{\sqrt{N}}\sum_{\mathbf r}e^{-i\mathbf k\cdot\mathbf r}c_{\mathbf r}.
\end{align}
With spin omitted, the Hamiltonian~\eqref{lattice_Hamiltonian} can thus be written in position space or in momentum space as
\begin{align}
H
&=
\sum_{\mathbf r,\mathbf r'}
\big[
t^{\phantom\dag}_{\mathbf r,\mathbf r'} c^{\dagger}_{\mathbf r}c^{\phantom\dag}_{\mathbf r'}
+ t^d_{\mathbf r,\mathbf r'} d^{\dagger}_{\mathbf r}d^{\phantom\dag}_{\mathbf r'}
+ (\Delta^{\phantom\dag}_{\mathbf r,\mathbf r'} c^{\dagger}_{\mathbf r}d^{\phantom\dag}_{\mathbf r'} + \rm{H.c.}) 
\big]
\label{real_space_ham}
\\
&=\sum_{\mathbf k} \big[
\xi^{\phantom\dag}_{\mathbf k} c^{\dagger}_{\mathbf k}c^{\phantom\dag}_{\mathbf k}
+ \xi^{d\phantom\dag}_{\mathbf k} d^{\dagger}_{\mathbf k}d^{\phantom\dag}_{\mathbf k}
+ (\Delta^{\phantom\dag}_{\mathbf k} c^{\dagger}_{\mathbf k}d^{\phantom\dag}_{\mathbf k} + \rm{H.c.})
\big].
\end{align}
The hoppings $t^{\phantom\dag}_{\mathbf r,\mathbf r'}$, $t^d_{\mathbf r,\mathbf r'}$ and coupling $\Delta_{\mathbf r,\mathbf r'}$ are the Fourier transforms of $\xi^{\phantom\dag}_{\mathbf k}$, $\xi^{d\phantom\dag}_{\mathbf k}$, and $\Delta^{\phantom\dag}_{\mathbf k}$, respectively.

The Hamiltonian can also be written as
\begin{align}
H&=\sum_{\mathbf k}
\begin{pmatrix}
c^\dagger_{\mathbf k}
&d^\dagger_{\mathbf k}
\end{pmatrix}
\begin{pmatrix}
\xi_{\mathbf k} & \Delta_{\mathbf k} \\
\Delta^*_{\mathbf k} &\xi^d_{\mathbf k}
\end{pmatrix}
\begin{pmatrix}
c_{\mathbf k}
\\d_{\mathbf k}
\end{pmatrix},
\label{af_Hamiltonian}
\end{align}
from which we can write a matrix Green's function
\begin{align}
\mathbf{G}_{\mathbf k}(z)
&=
\begin{pmatrix}
z-\xi_{\mathbf k} & -\Delta_{\mathbf k} \\
-\Delta^*_{\mathbf k} &z-\xi^d_{\mathbf k}
\end{pmatrix}^{-1}
\\
&=
\begin{pmatrix}
\dfrac{1}{z-\xi_{\mathbf k} 
- \dfrac{|\Delta_{\mathbf k}|^{2}}
{z-\xi^d_{\mathbf k}}}
&
\dfrac{\Delta_{\mathbf k}}{(z-\xi_{\mathbf k})(z-\xi^d_{\mathbf k}) - |\Delta_{\mathbf k}|^{2}}
\\
\dfrac{\Delta^*_{\mathbf k}}{(z-\xi_{\mathbf k})(z-\xi^d_{\mathbf k}) - |\Delta_{\mathbf k}|^{2}}
& \dfrac{1}{z-\xi^d_{\mathbf k} 
- \dfrac{|\Delta_{\mathbf k}|^{2}}
{z-\xi_{\mathbf k}}}
\end{pmatrix},
\end{align}
where $z$ is used as a shorthand for $\omega+i\eta$ with $\eta\rightarrow0$. The first element of this matrix is the lattice Green's function associated with the $c_{\mathbf k}$ operator,
\begin{align}
G(\mathbf k,\omega)
=
\dfrac{1}{\omega+i\eta-\xi_{\mathbf k} 
- \dfrac{|\Delta_{\mathbf k}|^{2}}
{\omega-\xi^d_{\mathbf k}}},
\label{target}
\end{align}
which is the time Fourier transform of $G^{R}(\mathbf k, t) = -i\langle\{c^{\phantom{\dag}}_{\mathbf k}(t),c^{\dag}_{\mathbf k}\}\rangle\theta(t)$. The last term in the denominator can be interpreted as a $\mathbf k$-dependent self-energy:
\begin{align}
\Sigma(\mathbf k,\omega) =
\dfrac{|\Delta_{\mathbf k}|^{2}}{\omega-\xi^d_{\mathbf k}}.
\label{target_self}
\end{align}



\paragraph*{Superlattice}

In quantum cluster methods~\cite{maier_quantum_2005,senechal_introduction_2008,senechal_quantum_2015}, another way to write Hamiltonian~\eqref{lattice_Hamiltonian} is to represent it on a superlattice of clusters, using the Fourier transform
\begin{align}
c_{\mathbf r} = c_{\tilde{\mathbf r}+\mathbf R} &=
\sqrt{\frac{N_c}{N}}
\sum_{\tilde{\mathbf k}}e^{i\tilde{\mathbf k}\cdot\tilde{\mathbf r}}
c_{\mathbf R}(\tilde{\mathbf k})
\label{zak_transform}
\end{align}
Here, $N_c$ is the number of sites in each cluster, and the position $\mathbf r$ of a lattice site is expressed as $\mathbf r = \tilde{\mathbf r} + \mathbf R$, where $\tilde{\mathbf r}$ is the position of the cluster (vector of the superlattice) and $\mathbf R$ is the position of the site within the cluster.
The corresponding separation in momentum space is $\mathbf k = \tilde{\mathbf k} + \mathbf K$, defined to satisfy the condition $e^{\mathbf K \cdot \tilde{\mathbf r}}=1$.
In this basis, Hamiltonian~\eqref{real_space_ham} becomes
\begin{align}
H
&= 
\sum_{\tilde{\mathbf k}} 
\sum_{\mathbf R \mathbf R^\prime} 
\begin{pmatrix}
c^\dag_{\mathbf R} (\tilde{\mathbf k})
&
d^\dag_{\mathbf R} (\tilde{\mathbf k})
\end{pmatrix}
\begin{pmatrix}
t_{\mathbf R \mathbf R^\prime}(\tilde{\mathbf k}) & \Delta_{\mathbf R \mathbf R^\prime} (\tilde{\mathbf k})
\\
\Delta^*_{\mathbf R \mathbf R^\prime} (\tilde{\mathbf k}) & t^d_{\mathbf R \mathbf R^\prime}(\tilde{\mathbf k})
\end{pmatrix}
\begin{pmatrix}
c^{\phantom{\dag}}_{\mathbf R^\prime} (\tilde{\mathbf k})
\\
d^{\phantom{\dag}}_{\mathbf R^\prime} (\tilde{\mathbf k})
\end{pmatrix}
\label{eq:ham2}
\\
&\equiv
\sum_{\tilde{\mathbf k}} 
\begin{pmatrix}
\mathbf c^\dag(\tilde{\mathbf k})
&\mathbf d^{\dag}(\tilde{\mathbf k})
\end{pmatrix}
\underbrace{
\begin{pmatrix}
\mathbf t(\tilde{\mathbf k}) & \mathbf \Delta(\tilde{\mathbf k})\\
\mathbf \Delta^{\dagger}(\tilde{\mathbf k}) &\mathbf t^{d}(\tilde{\mathbf k})
\end{pmatrix}
}_{\equiv \mathbf H (\tilde{\mathbf k})}
\begin{pmatrix}
\mathbf c(\tilde{\mathbf k}) \\
\mathbf d(\tilde{\mathbf k}) 
\end{pmatrix}.
\label{af_Hamiltonian3}
\end{align}
where the sum on $\tilde{\mathbf k}$ spans the Brillouin zone associated with the super-lattice. Note that $\mathbf c(\tilde{\mathbf k})$ and $\mathbf d(\tilde{\mathbf k})$ define spinors of length $N_c$. Thus, the $2\times2$ Hamiltonian matrix above is in fact of dimension $2N_c\times 2N_c$,
\begin{align}
\mathbf H (\tilde{\mathbf k})
=
\begin{pmatrix}
\mathbf t(\tilde{\mathbf k}) & \mathbf \Delta(\tilde{\mathbf k})\\
\mathbf \Delta^{\dagger}(\tilde{\mathbf k}) &\mathbf t^{d}(\tilde{\mathbf k})
\end{pmatrix}
\end{align}
The elements of the hopping blocks $\mathbf t(\tilde{\mathbf k})$, $\mathbf t^{d}(\tilde{\mathbf k})$ and gap block $\mathbf \Delta(\tilde{\mathbf k})$ can be obtained as
\begin{align}
t_{\mathbf R, \mathbf R'} (\tilde{\mathbf k})
&=
\sum_{\tilde{\mathbf r}} e^{-i\tilde{\mathbf k}\cdot\tilde{\mathbf r}} t_{\tilde{\mathbf r} + \mathbf R, \mathbf R'}
\\
t^{d}_{\mathbf R, \mathbf R'} (\tilde{\mathbf k})
&=
\sum_{\tilde{\mathbf r}} e^{-i\tilde{\mathbf k}\cdot\tilde{\mathbf r}} t^{d}_{\tilde{\mathbf r} + \mathbf R, \mathbf R'}
\label{superlattice_h}
\\
\Delta_{\mathbf R, \mathbf R'} (\tilde{\mathbf k})
&=
\sum_{\tilde{\mathbf r}} e^{-i\tilde{\mathbf k}\cdot\tilde{\mathbf r}} \Delta_{\tilde{\mathbf r} + \mathbf R, \mathbf R'}
\end{align}
with the sum over $\tilde{\mathbf r}$ spanning all vectors of the super-lattice.
In this basis, the lattice Green's function $\mathbf G(\tilde{\mathbf k},\omega)$ associated with the $\mathbf c(\tilde{\mathbf k})$ spinor (the Fourier transform of $\mathbf G^{R}(\tilde{\mathbf k}, t) = -i\langle\{\mathbf c(\tilde{\mathbf k}, t),\mathbf c^{\dag}(\tilde{\mathbf k})\}\rangle\theta(t)$) is obtained by taking the Schur complement of the block $[z-\mathbf t^{d}(\tilde{\mathbf k})]$ from the matrix $[z-\mathbf H (\tilde{\mathbf k})]$:
\begin{align}
\mathbf G(\tilde{\mathbf k},\omega)
&=
\frac{1}{
\omega+i\eta - \mathbf t (\tilde{\mathbf k})
- \mathbf \Delta^\dag (\tilde{\mathbf k}) \dfrac{1}{\omega-\mathbf t^d(\tilde{\mathbf k})} \mathbf \Delta (\tilde{\mathbf k})
}.
\label{eq:Green1}
\end{align}
This yields a matrix representation of the lattice self-energy
\begin{align}
\mathbf \Sigma(\tilde{\mathbf k},\omega)
&= 
\mathbf \Delta^\dag (\tilde{\mathbf k}) \dfrac{1}{\omega-\mathbf t^d(\tilde{\mathbf k})} \mathbf \Delta (\tilde{\mathbf k}).
\end{align}

\paragraph*{Single cluster}
When restricting the above Hamiltonian to a single cluster, all matrices loose their dependence on $\tilde{\mathbf k}$. Formally, this corresponds to taking only the cluster at $\tilde{\mathbf r}=0$, which is equivalent to
\begin{align}
\mathbf H_c = \frac{N_c}{N}\sum_{\tilde{\mathbf k}}\mathbf H (\tilde{\mathbf k}).
\label{coarse_grain}
\end{align}
Starting from that cluster Hamiltonian $\mathbf H_c$, the same steps as above lead to the cluster Green's function
\begin{align}
\mathbf G_c(\omega)
&=
\frac{1}{
\omega+i\eta - \mathbf t_c - \mathbf \Delta_c^\dag \dfrac{1}{z-\mathbf t_c^d} \mathbf \Delta_c
},
\end{align}
with cluster self-energy
\begin{align}
\mathbf \Sigma_c(\omega)
&= 
\mathbf \Delta_c^\dag \dfrac{1}{\omega-\mathbf t_c^d} \mathbf \Delta_c.
\end{align}
which is equation~\eqref{eq:sigma1} in the main text.
Note that although $\mathbf t_c$, $\mathbf t^d_c$, and $\mathbf \Delta_c$ can all be obtained as
\begin{align}
\mathbf t_c &= \frac{N_c}{N}\sum_{\tilde{\mathbf k}}\mathbf t (\tilde{\mathbf k}),
\\
\mathbf t^d_c &= \frac{N_c}{N}\sum_{\tilde{\mathbf k}}\mathbf t^d (\tilde{\mathbf k}),
\label{cluster_h}
\\
\mathbf \Delta_c &= \frac{N_c}{N}\sum_{\tilde{\mathbf k}}\mathbf \Delta (\tilde{\mathbf k}),
\end{align}
this is not the case for $\mathbf G_c(\omega)$ and $\mathbf \Sigma_c(\omega)$,
\begin{align}
\mathbf G_c(\omega) &\neq \frac{N_c}{N}\sum_{\tilde{\mathbf k}}\mathbf G(\tilde{\mathbf k},\omega),
\label{not_green}
\\
\mathbf \Sigma_c(\omega) &\neq \frac{N_c}{N}\sum_{\tilde{\mathbf k}}\mathbf \Sigma(\tilde{\mathbf k},\omega).
\label{not_self}
\end{align}
This is one of the fundamental reasons why the standard periodization schemes cannot recover the original lattice Green's function~\eqref{eq:Green1}, as explained in what follows.

\paragraph*{Compact tiling}
Periodization tries to reverse the ``clusterization'' or ``coarse-graining'' realized in~\eqref{coarse_grain}.
Since the latter is applied to the Hamiltonian, it is the Hamiltonian which should be reconstructed, not the Green's function, self-energy, or cumulant.
Furthermore, to reverse~\eqref{coarse_grain}, one cannot simply Fourier transform back from representation~\eqref{zak_transform} as done in the M- and G-schemes.
One must also rebuild the lost inter-cluster elements
\begin{align}
\delta \mathbf t(\tilde{\mathbf k}) &= \mathbf t (\tilde{\mathbf k}) - \mathbf t_c,
\\
\delta \mathbf t^d(\tilde{\mathbf k}) &= \mathbf t^d (\tilde{\mathbf k}) - \mathbf t^d_c,
\\
\delta \mathbf \Delta(\tilde{\mathbf k}) &= \mathbf \Delta (\tilde{\mathbf k}) - \mathbf \Delta_c.
\end{align}
To do so, we propose the compact tiling procedure illustrated at Fig.~\eqref{fig_compact_tiling} for the $2\times 2$ case. Let us now justify this procedure.

In general, reversing a sum like~\eqref{coarse_grain} is a fundamentally ill-defined task.
However, with the hypothesis of translation invariance, the sum in~\eqref{coarse_grain} removes information in a very structured way, as we can see, for example, by subtituting~\eqref{superlattice_h} in~\eqref{cluster_h} 
\begin{align}
\label{compact_first}
[\mathbf t^d_c]_{\mathbf R, \mathbf R'} 
&= 
\frac{N_c}{N}\sum_{\tilde{\mathbf k}} 
t^d_{\mathbf R, \mathbf R'}(\tilde{\mathbf k})
\\
&= 
\frac{N_c}{N}\sum_{\tilde{\mathbf k}} 
\sum_{\tilde{\mathbf r}}
e^{-i\tilde{\mathbf k}\cdot\tilde{\mathbf r}}
t^d_{\tilde{\mathbf r} + \mathbf R, \mathbf R'}
\\
&=
\sum_{\tilde{\mathbf r}} 
\left(
\frac{N_c}{N}\sum_{\tilde{\mathbf k}} 
e^{-i\tilde{\mathbf k}\cdot\tilde{\mathbf r}}
\right)
t^d_{\tilde{\mathbf r} + \mathbf R, \mathbf R'}
\\
&=
\sum_{\tilde{\mathbf r}} 
\delta_{\tilde{\mathbf r}=0}
t^d_{\tilde{\mathbf r} + \mathbf R, \mathbf R'}
\\&
=
t^d_{\mathbf R, \mathbf R'}.
\label{compact_last}
\end{align}
The above equations relate the components of the cluster operator~$\mathbf t^d_c$ to the lattice~$t^d_{\mathbf r, \mathbf r'}$. Our procedure of compact tiling can be seen as rewinding this sequence, starting from the cluster operator~\eqref{compact_last}, and adding back the phase factors to rebuild the lattice operator~$t^d_{\mathbf R, \mathbf R'}(\tilde{\mathbf k})$ under the sum in~\eqref{compact_first}. Components not contained within one cluster cannot be reconstructed (\emph{e.g.} $t''$ is lost in the $2\times2$ case). Note that since the equivalent of~\eqref{compact_first}-\eqref{compact_last} also relates $\mathbf \Delta_c$ to the lattice $\Delta_{\mathbf r, \mathbf r'}$, we can use compact tiling to reconstruct~$\Delta_{\mathbf R, \mathbf R'}(\tilde{\mathbf k})$ as well. Clearly, compact tiling works best when hoppings are short-ranged.

However, if we used compact tiling in an attempt to rebuild
\begin{align}
\delta \mathbf G(\tilde{\mathbf k},\omega) &= \mathbf G (\tilde{\mathbf k}) - \mathbf G_c(\omega),
\\
\delta \mathbf \Sigma(\tilde{\mathbf k},\omega) &= \mathbf \Sigma (\tilde{\mathbf k}) - \mathbf \Sigma_c(\omega),
\end{align}
it would fail because of~\eqref{not_green} and~\eqref{not_self}. We suspect that similar procedures suggested in earlier works for the self-energy and the cumulant~\cite{biroli_cluster_2004,stanescu_strong_2004} also failed because of~\eqref{not_green} and~\eqref{not_self}.

\paragraph*{$2\times2$ case}
The specific example considered in the main text is the $2\times2$ cluster, a geometry often studied with c-DMFT. In this case, the spinor reprepresentation is:
\begin{align}
\mathbf c^\dag_{\tilde{\mathbf k}} =
\begin{pmatrix}
c^\dag_{\mathbf R_1}(\tilde{\mathbf k}) &
c^\dag_{\mathbf R_2}(\tilde{\mathbf k}) &
c^\dag_{\mathbf R_3}(\tilde{\mathbf k}) &
c^\dag_{\mathbf R_4}(\tilde{\mathbf k})
\end{pmatrix},
\end{align}
where the $\mathbf R_i$ are the four positions of the cluster, enumerated from 1 to 4, counter-clockwise.
In our example, we used a local coupling only,
\begin{align}
\mathbf \Delta (\tilde{\mathbf k})
=
\mathbf \Delta_c
&=
\begin{pmatrix}
\Delta&0 &0 &0   \\
0&\Delta&0 &0   \\
0&0  &\Delta&0   \\
0&0  &0 &\Delta
\end{pmatrix},
\end{align}
The dispersion $\mathbf t(\tilde{\mathbf k})$ can be written as $\mathbf t(\tilde{\mathbf k})=\mathbf t_c+\delta\mathbf t(\tilde{\mathbf k})$ with intra-cluster hoppings:
\begin{align}
\mathbf t_c
&=
-
\begin{pmatrix}
 \mu\phantom{'}  &  t\phantom{'}  &  t'            &  t\phantom{'}  \\
 t\phantom{'}     &  \mu\phantom{'}  &  t\phantom{'}  &  t' \\
 t'               &  t\phantom{'}     &  \mu\phantom{'}  &  t\phantom{'}  \\
 t\phantom{'}     &  t'               &  t\phantom{'}     &  \mu\phantom{'}
\end{pmatrix},
\label{tc}
\end{align}
and inter-cluster hoppings: 
\begin{widetext}
\begin{align}
\delta \mathbf t(\tilde{\mathbf k}) = -
\begin{pmatrix}
2t^{\prime\prime}(\cos(2\tilde{k}_x)+\cos(2\tilde{k}_y)) & t e^{-2i\tilde{k}_x} & t^{\prime} (e^{-2i\tilde{k}_x} + e^{-2i\tilde{k}_y} + e^{-2i(\tilde{k}_x+\tilde{k}_y)}) & t e^{-2i\tilde{k}_y} \\
t e^{2i\tilde{k}_x} & 2t^{\prime\prime}(\cos(2\tilde{k}_x)+\cos(2\tilde{k}_y)) & t e^{-2i\tilde{k}_y} &  t^{\prime} (e^{2i\tilde{k}_x} + e^{-2i\tilde{k}_y} + e^{2i(\tilde{k}_x-\tilde{k}_y)}) \\
t^{\prime} (e^{2i\tilde{k}_x} + e^{2i\tilde{k}_y} + e^{2i(\tilde{k}_x+\tilde{k}_y)})  & t e^{2i\tilde{k}_y} & 2t^{\prime\prime}(\cos(2\tilde{k}_x)+\cos(2\tilde{k}_y)) & t e^{2i\tilde{k}_x} \\
t e^{2i\tilde{k}_y} &  t^{\prime} (e^{-2i\tilde{k}_x} + e^{2i\tilde{k}_y} + e^{2i(-\tilde{k}_x+\tilde{k}_y)})  & t e^{-2i\tilde{k}_x} & 2t^{\prime\prime}(\cos(2\tilde{k}_x)+\cos(2\tilde{k}_y))
\end{pmatrix}.
\label{deltatktilde}
\end{align}
\end{widetext}
The auxiliary dispersion $\xi^{d}_{\mathbf k}=\xi_{\mathbf k + \mathbf Q}$ with $\mathbf Q=(\pi,\pi)$ leads to matrices $\mathbf t^d_c$ and $\delta \mathbf t^d(\tilde{\mathbf k})$ identical to the above with $t$ replaced by $-t$, as seen for $\mathbf t^d_c$ at equation~\eqref{tdc}. By observation of~\eqref{tc} and~\eqref{deltatktilde} we see that $\delta \mathbf t(\tilde{\mathbf k})$ can be rebuilt from $\mathbf t_c$, except for the term with $t''$, as discussed in the main text. This "reconstruction by observation" correspond to the compact tiling scheme suggested at Fig.~\ref{fig_compact_tiling}.

\end{document}